\documentclass [a4paper,11pt]{article}
\pdfoutput=1
\usepackage{amsmath}
\usepackage{amsfonts}
\usepackage{amssymb}
\usepackage{graphicx}
\usepackage{xcolor}
\usepackage{ulem}
\usepackage{graphicx}
\usepackage[body={6.5in, 9in}, right=1in, top=1in]{geometry}
\usepackage{amssymb}
\usepackage{amsmath} 
\usepackage[numbers,sort&compress]{natbib}

\makeatletter
\def\section{\@startsection {section}{1}{\z@}{-3.5ex plus -1ex minus
 -.2ex}{2.3ex plus .2ex}{\large\bf}}
\def\subsection{\@startsection{subsection}{2}{\z@}{-3.25ex plus -1ex
minus -.2ex}{1.5ex plus .2ex}{\normalsize\bf}}
\makeatother
\makeatletter

\@addtoreset{equation}{section}

\makeatother
\def\be{\begin{equation}}
\def\ee{\end{equation}}

\newcommand\eea{\end{eqnarray}}
\newcommand\bea{\begin{eqnarray}}

\newcommand{\rmd}{\textrm{d}}

\def\({\left(}
\def\){\right)}

\usepackage{soul}
\newcommand{\Comment}[1]{{}}
\definecolor{MyDarkBlue}{rgb}{0.15,0.15,0.45}
\usepackage[linktocpage=true]{hyperref}
\hypersetup{
colorlinks=true,
citecolor=MyDarkBlue,
linkcolor=MyDarkBlue,
urlcolor=MyDarkBlue,
}

\begin{document}
\def\thefootnote{\fnsymbol{footnote}}

\begin{center}
\Large{\textbf{Inequivalence of Coset Constructions  \\ [0.1cm]for Spacetime Symmetries}} \\[0.5cm]
 
\large{Paolo Creminelli$^{\rm a,b}$, Marco Serone$^{\rm c,a,d}$, Gabriele Trevisan$^{\rm c,d}$ and Enrico Trincherini$^{\rm e,f}$}
\\[0.5cm]

\small{
\textit{$^{\rm a}$ Abdus Salam International Centre for Theoretical Physics\\ Strada Costiera 11, 34151, Trieste, Italy}}

\vspace{.2cm}

\small{
\textit{$^{\rm b}$ Institute for Advanced Study, Princeton, New Jersey 08540, USA}}

\vspace{.2cm}

\small{
\textit{$^{\rm c}$ SISSA, via Bonomea 265, 34136, Trieste, Italy}}

\vspace{.2cm}

\small{
\textit{$^{\rm d}$ INFN - Sezione di Trieste, 34151 Trieste, Italy}}

\vspace{.2cm}

\small{
\textit{$^{\rm e}$ Scuola Normale Superiore, piazza dei Cavalieri 7, 56126, Pisa, Italy}}

\vspace{.2cm}

\small{
\textit{$^{\rm f}$ INFN - Sezione di Pisa, 56100 Pisa, Italy}}

\vspace{.2cm}

\end{center}

\vspace{.8cm}

\hrule \vspace{0.3cm}
\noindent \small{\textbf{Abstract}\\
Non-linear realizations of spacetime symmetries can be obtained by a generalization of the coset construction valid for internal ones.
The physical equivalence of different representations for spacetime symmetries is not obvious, since their relation involves not only a redefinition of the fields but also a field-dependent change of coordinates. A simple and relevant spacetime symmetry is obtained by the contraction of the 4D conformal group that leads to the Galileon group.
We analyze two non-linear realizations of this group, focusing in particular on the propagation of signals around non-trivial backgrounds. The aperture of the lightcone is in general different in the two representations and in particular a free (luminal) massless scalar is mapped in a Galileon theory which admits superluminal propagation. We show that in this theory, if we consider backgrounds that vanish at infinity, there is no asymptotic effect: the displacement of the trajectory integrates to zero, as can be expected since the S-matrix is trivial. Regarding local measurements, we show that the puzzle is solved taking into account that a local coupling with fixed sources in one theory is mapped into a non-local coupling and we show that this effect compensates the different lightcone. Therefore the two theories have a different notion of locality. The same applies to the different non-linear realizations of the conformal group and we study the particular case of a cosmologically interesting background: the Galilean Genesis scenarios.
} 
\vspace{0.3cm}
\noindent
\hrule
\def\thefootnote{\arabic{footnote}}
\setcounter{footnote}{0}

\section{Introduction}

The treatment of non-linear realizations of internal symmetries in quantum field theory is by now well understood. If a symmetry group $G$ is spontaneously broken to a subgroup $H$, the Goldstone theorem ensures that one massless spin zero particle occurs for each broken symmetry generator. Remarkably, the low-energy dynamics of Goldstone bosons is quite insensitive to the specific UV theory that originated them and is essentially governed by the group theory structure of the 
coset space $G/H$, as originally shown by Callan, Coleman, Wess and Zumino (CCWZ)  \cite{Coleman:1969sm,Callan:1969sn}. For a given coset $G/H$, 
there is an infinite number of possible parameterizations leading to different Lagrangians, but these are all physically equivalent, since local field redefinitions do not change physical observables.

The extension of such considerations to spacetime symmetries is far from trivial. First of all, the Goldstone theorem no longer applies and we lose the one-to-one correspondence between massless spin zero particles and broken generators. 
The number of Goldstone particles is less than the number of broken generators because certain constraints among the Goldstones can be imposed, the so-called ``inverse Higgs phenomenon" \cite{Ivanov:1975zq}.\footnote{
See \cite{Endlich:2013vfa,Brauner:2014aha} for recent analyses on the meaning of the inverse Higgs phenomenon.} A notable example is the spontaneous breaking of conformal symmetry down to the Poincar\'e group \cite{Salam:1970qk,Isham:1971dv}. In four dimensions the conformal group is isomorphic to SO(4,2) but of the naively expected 5 Goldstones
only one is necessarily present, the dilaton. 

The technical generalization of the CCWZ coset construction to spacetime symmetries has been developed in \cite{Volkov:1973vd,Ivanov:1975zq}, but the physical interpretation of this construction is
perhaps not totally understood. The spacetime coordinates enter now explicitly in the definition of the element of the coset.  
There is still in principle an infinite number of ways of parametrizing the relevant coset but finding explicit tractable parametrizations is not straightforward.
Moreover, and most importantly for the present work, it is not clear whether and in what sense they are physically equivalent or not. 

Two relevant non-linear realizations  of the conformal group (and their relation), dubbed DBI and Weyl representations in \cite{Creminelli:2013fxa}, have been derived in \cite{Bellucci:2002ji} from the coset construction of \cite{Volkov:1973vd,Ivanov:1975zq}. They are related by a complicated field and spacetime coordinate redefinition.
We have recently shown that the relations between such representations can be seen as a change of coordinates in AdS$_5$ and lead to the same $2\rightarrow 2$ dilaton scattering amplitude \cite{Creminelli:2013fxa}. However, the complete physical equivalence of the two representations was not clear, since around non-trivial backgrounds the aperture of the lightcone of the dilaton in the two representations is different and theories which have only subluminal propagation are mapped into theories which allow superluminality. 

With this motivation, the aim of this paper is to shed light on the general question of whether different parametrizations of cosets involving spacetime symmetries are physically equivalent or not.

We start in section \ref{sec:TwistAlgebra} by introducing what is going to be our prominent example of coset construction: Gal$(3+1,1)/$ISO$(3,1)$. This coset,  introduced in  \cite{Goon:2012dy} (see also \cite{Kamimura:2013mia}) to study 
Galileons \cite{Nicolis:2008in}, can be seen as the contraction of the conformal coset SO$(4,2)/$ISO$(3,1)$. It has the advantage, with respect to the latter, of being simpler to treat, yet keeping the same qualitative features. 
The two non-linear realizations of this coset that we consider are contractions of the DBI and Weyl representations
of \cite{Bellucci:2002ji,Creminelli:2013fxa}. They are related to each other by a simple-looking field redefinition, eq.~(\ref{eq.coordtransf0}), that  crucially involves the spacetime coordinates as well. Written as an ordinary field redefinition (i.e. keeping spacetime coordinates fixed), the field redefinition (\ref{eq.coordtransf0}) can be expanded in an infinite series of higher derivative terms.

Recently the same field redefinition has also been found directly in the case of the standard (i.e. not conformal) Galileons in \cite{deRham:2013hsa} (see also \cite{Curtright:2012gx}). In fact for the coset Gal$(3+1,1)/$ISO$(3,1)$, since both the DBI and the Weyl operators reduce to the same set of five standard Galileons,  
the transformation that connects the two representations simply maps the five operators into themselves.    
In particular this gives rise to the apparent paradox presented in \cite{deRham:2013hsa}: a free massless scalar in one representation is mapped in the other into a non trivial Galileon theory in which superluminality seems to occur! In section \ref{sec.superluminal} we address this puzzle working at first order in the amplitude of the background. First of all we show that, in the theory obtained from the free theory by a field redefinition, there is no asymptotic effect  if one considers a localized background which vanishes at infinity. The trajectory deviates from the lightcone, but the net effect integrates to zero asymptotically. This squares with the fact that the two theories have the same (trivial!) S-matrix, so that they should give the same answer to any question which involves asymptotic states.

We then proceed to consider what happens if one measures the superluminality locally within a background, coupling the field to an external source, which does not transform when one changes the coset parametrization. The apparent paradox is solved by realizing  that a local coupling to a source in one representation becomes non-local in the other representation. We analyze in detail how the non-local coupling provides a delay in the signal propagation  that precisely compensates for the apparent superluminal behaviour. We thus conclude that the two representations have a different notion of locality and thus are  not physically equivalent.

In section \ref{sec.more} we provide several additional calculations to support this conclusion. In particular we show that there are no asymptotic effects at second order for a generic background in subsec.~\ref{sub.secondorder}. Furthermore we consider a cylindrically symmetric background in subsec. \ref{sub.cylinder}, and these two examples illustrate that the cancellation of asymptotic effects happens only for the particular linear combination of the five Galileons obtained by mapping the free theory. In subsec.~\ref{sub.sources} we show that our arguments apply even when sources are included. Then in subsec.~\ref{sub.DGP} we reconsider the Dvali-Gabadadze-Porrati (DGP) model \cite{Dvali:2000hr} and we verify we have asymptotic superlumnality in the case of a spherically symmetric solution.
Finally, in subsec.~\ref{subsec:allorders}, we come back to our main example and show how luminality is recovered to all orders in the background amplitude for a specific, translationally invariant, classical configuration when non-local couplings are considered.
 
In section \ref{sec:conformal} we show how
the same mechanism applies to the conformal coset and explains the different aperture of the dilaton lightcone found in  \cite{Creminelli:2013fxa} in the context of the Genesis scenario \cite{Creminelli:2010ba,Hinterbichler:2012yn}. In section \ref{sec:con} we summarize our findings and conclude with the general lesson that can be learned from these examples. 
For completeness, we report in appendix \ref{app:TdR} how the Galileon map found in \cite{deRham:2013hsa}
is recovered through our coset construction.
In appendix \ref{app:finite} we explain why the field redefinition we consider (a sum of an infinite number of higher derivative terms) cannot be truncated;
if this is done either the superluminal effect is not measurable or we are forced to study higher derivative equations of motion. 

\section{Non-linear Realizations of the Galileon Group}

\label{sec:TwistAlgebra}

The existence of a mapping between operators invariant under the non-linearly realized Galileon symmetry $\pi(y) \to \pi(y) + c + b_\mu y^\mu $ can be traced back to the existence of two realizations related by a ``twist" of the same symmetry breaking pattern Gal$(3+1,1)/$ISO$(3,1)$.
The Galileon algebra Gal$(3+1,1)$ \cite{Goon:2012dy} can be written as
\begin{equation}\label{galalgebra1}
\begin{split}
& \left[M_{\rho \sigma},B_{\nu }\right]=\eta_{\nu \rho  }B_{\sigma}-\eta_{\nu \sigma  }B_{\rho}\ , \\
&\left [P_\mu,B_\nu\right]=\eta_{\mu\nu}C\ ,\\
&\left [C,P_\mu\right]=0\ ,\\
&\left [C,B_{\mu}\right]=0\ .\\
\end{split}
\end{equation}
Going through the coset construction for non-linearly realized spacetime symmetries it is easy to write the covariant Cartan 1-form in terms of the generators of the algebra (\ref{galalgebra1}):
\begin{equation}
\begin{split}
g^{-1} \text{d}g&=e^{-\Omega^\mu B_\mu}e^{-\pi C}e^{-y^\mu P_\mu}e^{(y^\mu+\text{d}y^\mu) P_\mu}e^{(\pi+\text{d}\pi) C}e^{(\Omega^\mu+\text{d}\Omega^\mu) B_\mu}\\
&=\text{d}y^\mu P_\mu+( \text{d}y^\mu \Omega_\mu + \text{d}\pi) C+\text{d}\Omega^\mu B_\mu \ .
\end{split}
\end{equation}
The field $\Omega_\mu$ can be fixed in terms of the only physical degree of freedom $\pi$ by imposing the inverse Higgs constraint 
\begin{equation}
\begin{split}
\Omega_\mu(y)=-\frac{\partial \pi}{\partial y^\mu}  \ .
\end{split}
\end{equation}
Now consider the following linear combination of generators
\begin{equation}
\begin{split}
\hat B_\mu = \frac{1}{\sqrt{2}L}B_\mu+\frac{L}{\sqrt{2}} P_\mu \, , \qquad \hat C= \frac{1}{\sqrt{2}L} C\ .
\end{split}
\end{equation}
It is immediate to verify that the twisted generators  $(\hat B_\mu, \hat C)$ together with $(M_{\mu\nu}, P_\mu)$ also satisfy the Galileon algebra in the same form of eq.~(\ref{galalgebra1}).
The Cartan 1-form expanded in terms of these generators with a new set of spacetime coordinates $x^\mu$ and Goldstone bosons $q(x)$, $\Lambda_\mu(x)$ can then be written as  
\begin{equation}
\begin{split}
g^{-1} \text{d}g&=e^{-\Lambda^\mu \hat B_\mu}e^{-q \hat C}e^{-x^\mu P_\mu}e^{(x^\mu+\text{d}x^\mu) P_\mu}e^{(q+\text{d}q) \hat C}e^{(\Lambda^\mu+\text{d}\Lambda^\mu) \hat B_\mu}\\
&=\text{d}x^\mu P_\mu+( \text{d}x^\mu \Lambda_\mu + \text{d}q) \hat C+\text{d}\Lambda^\mu \hat B_\mu \\
&=\Big( \text{d}x^\mu+ \frac{L}{\sqrt{2}}  \text{d}\Lambda^\mu \Big) P_\mu+( \text{d}x^\mu \Lambda_\mu + \text{d}q)\frac{1}{\sqrt{2}L} C+ \frac{1}{\sqrt{2}L}  \text{d}\Lambda^\mu B_\mu \ ,
\end{split}
\end{equation}
and the inverse Higgs constraint fixes $\Lambda_\mu$ to be
\begin{equation}
\begin{split}
\Lambda_\mu(x)=-\frac{\partial q}{\partial x^\mu}  \ .
\end{split}
\end{equation}
The mapping between the two coordinate systems and the Goldstone $\pi(y)$ and $q(x)$ is obtained by equating the two expressions for the Cartan form: 
\begin{equation}	\label{eq.coordtransf0}
\begin{split}
y^\mu&=x^\mu- L \partial^\mu q(x)\ ,\\
\frac{\partial \pi(y)}{\partial y^\mu} &= \frac{1}{L}\frac{\partial q(x)}{\partial x^\mu} \ ,\\
\pi(y)&= \frac{q(x)}{L}-\frac{1}{2} (\partial q(x))^2 \ ,
\end{split}
\end{equation}
where $q$ has been redefined as $\sqrt{2} q$ so that the first line, apart from a minus sign\footnote{The slight differences in the notation with respect to \cite{deRham:2013hsa} are discussed in appendix \ref{app:TdR}.}, is exactly the coordinate transformation of \cite{deRham:2013hsa}.
Since the two representations satisfy the {\it same} algebra, the invariant operators will be the same in both the $(\pi, y^\mu)$ and $(q, x^\mu)$ basis. 

It is clear that a generic operator written in terms of the $\pi$ field is generally mapped to a complicated infinite series of  higher derivative operators when re-expressed in terms of the field $q$ (and viceversa). 
Remarkably, there is a set of special operators that are mapped into themselves  under the mapping  (\ref{eq.coordtransf0}) \cite{deRham:2013hsa}.
These are the so called Galileon operators \cite{Nicolis:2008in}. In $D=4$ they are 5 operators, constructed with $n$ elementary fields and $2n-2$ derivatives $(n=1,...,5)$. They are the only Galileon invariant operators that give second order equations of motion.

The map between the Galileon operators can be obtained by the map found in \cite{Creminelli:2013fxa} between the conformal Galileons (i.e. Galileon operators that non-linearly realize the conformal group, where $\pi$ is identified with the dilaton).
The Galileon algebra in fact can be recovered taking an \.{I}n\"on\"u-Wigner group contraction of the conformal algebra $SO(4,2)$.
The relevant part of the latter reads
\begin{equation}
\begin{split}
& \left[M_{\rho \sigma},K_{\nu }\right]=\eta_{\nu\rho}K_{\sigma}-\eta_{\nu\sigma}K_{\rho}\ , \\
&\left [P_\mu,K_\nu\right]=2(-\eta_{\mu\nu}D+2M_{\mu\nu})\ ,\\
&\left [D,P_\mu\right]=P_\mu\ ,\\
&\left [D,K_{\mu}\right]=-K_\mu\ .\\
\end{split}
\end{equation}
This is the standard basis of the conformal algebra.   In this representation the operators of the coset SO$(4,2)/$ISO$(3,1)$ that give second order equations of motion corresponds to the 5 conformal Weyl Galileons (we follow the nomenclature of \cite{Creminelli:2013fxa}) written in terms of the Goldstone boson $\pi(y)$.

The group contraction of SO$(4,2)$ to the Galileon algebra (\ref{galalgebra1}) is then obtained by setting
\begin{equation}\label{dilation1}
K_\mu \rightarrow \epsilon^{-1} B_\mu \,,\qquad \qquad D \rightarrow -\frac{1}{2} \epsilon^{-1} C
\end{equation}
and taking the limit $\epsilon \rightarrow 0$. In this way the standard Galileon operators \cite{Nicolis:2008in} are recovered as the $\pi \to 0$ limit of the Weyl ones.  

The twist of the generators \cite{Bellucci:2002ji, Creminelli:2013fxa}
\begin{equation}
\begin{split}
\hat K_\mu = \frac{1}{\sqrt{2}L}K_\mu+\frac{L}{\sqrt{2}} P_\mu\,, \ \qquad \qquad \hat D= \frac{1}{\sqrt{2}L} D\ ,
\end{split}
\end{equation}
gives the conformal algebra in a different basis:
\begin{equation}
\begin{split}
& \left[\hat K_{\mu},\hat K_{\nu }\right]=4 M_{\mu\nu}\ , \\
& \left[M_{\rho \sigma},\hat K_{\nu }\right]=\eta_{\nu\rho  }\hat K_{\sigma}-\eta_{\nu\sigma  }\hat K_{\rho}\ , \\
&\left [P_\mu,\hat K_\nu\right]=2\Big(-\eta_{\mu\nu}\hat D+\frac{\sqrt{2}}{L}M_{\mu\nu}\Big)\ ,\\
&\left [\hat D,P_\mu\right]= \frac{1}{\sqrt{2}L}  P_\mu\ ,\\
&\left [\hat D,\hat K_{\mu}\right]=- \frac{1}{\sqrt{2}L}\hat K_\mu+ P_\mu\ .\\
\end{split}
\end{equation}
In this representation the five Galileon operators of the coset are the DBI Galileons \cite{deRham:2010eu}. Also in this case the group contraction is taken by setting
\begin{equation}\label{dilation2}
\hat K_\mu \rightarrow \epsilon^{-1}\hat  B_\mu\,, \qquad\qquad \hat D \rightarrow -\frac{1}{2}\epsilon^{-1} \hat C\ ,
\end{equation}
and in the limit $\epsilon \rightarrow 0$ gives the algebra (\ref{galalgebra1}). As a consequence, the DBI Galileons as well reduce in the $q\rightarrow 0$ limit
to the same five standard Galileons as before.
As shown in \cite{Creminelli:2013fxa}, in going from one representation of the conformal group to the other the two sets of five conformal Galileons (DBI and Weyl) are mapped one into the other. Since after the group contraction both reduce to the ordinary Galileons, the conformal mapping in the appropriate limit becomes precisely the duality of Galileon theories reported in \cite{deRham:2013hsa}.
The details of the limit, the explicit expression of the mapping and the comparison with \cite{deRham:2013hsa} are given in appendix \ref{app:TdR}. 

Since the mapping between the two representations is a central part of the apparent paradox that we are going to discuss in section \ref{sec.superluminal}, it is worth spending some few lines on it.
In order for $\pi$ and $q$ to have the same dimensions from now on we further redefine $\pi\rightarrow L \pi $, $q\rightarrow  L^2 q $ and denote by $L=\Lambda^{-1}$; the map then reads
\begin{equation}	\label{eq.coordtransf}
\begin{split}
y^\mu&=x^\mu- \frac{1}{\Lambda^3} \partial^\mu q(x)\ ,\\
\frac{\partial \pi(y)}{\partial y^\mu} &= \frac{\partial q(x)}{\partial x^\mu} \ ,\\
\pi(y)&= q(x)-\frac{1}{2\Lambda^3} (\partial q(x))^2 \,.
\end{split}
\end{equation}
These three conditions are redundant and one can specify only two of them. For example, given the change of coordinates and the relation between the derivatives of the fields, one can infer the field redefinition by taking the derivative of the ansatz
\begin{equation}
\begin{split}
\pi (y) = q(x)+ f(\partial q) \ .
\end{split}
\end{equation}
Since this is short but instructive we will do it explicitly. One finds
\begin{equation}
\begin{split}
\frac{\partial}{\partial x^\mu} q(x)=\frac{\partial}{\partial y^\mu} \pi(y)= \frac{\partial x^\nu }{\partial y^\mu}\frac{\partial}{\partial x^\nu}\left( q(x)+ f(\partial q) \right)\Rightarrow f(\partial q)= -\frac{1}{2\Lambda^3}(\partial q)^2 \ ,
\end{split}
\end{equation} 
under the condition of invertibility of the matrix $J= \frac{\partial x }{\partial y}$, which is the Jacobian of the change of coordinates. Viceversa, given the coordinate transformation and the field redefinition, by taking the derivative of $\pi$ one finds
\begin{equation}
\begin{split}
\frac{\partial}{\partial y^\mu} \pi(y)= \frac{\partial x^\nu }{\partial y^\mu}\frac{\partial}{\partial x^\nu}\left( q(x) -\frac{1}{2\Lambda^3}(\partial q)^2 \right)\Rightarrow \frac{\partial}{\partial y^\mu} \pi(y)=\frac{\partial}{\partial x^\mu} q(x) \ ,
\end{split}
\end{equation} 
and thus the derivatives are related as expected from the second line of eq.~(\ref{eq.coordtransf}), again under the condition of invertibility of the  Jacobian matrix. 
Of course, in order to retain the equivalence of the two theories, the minimal assumption that we have to make is the invertibility of the map between them. From the field redefinition alone, it is not easy to guess what is $q$ as a function of $\pi$, since this would involve solving a nonlinear differential equation, but the task is made completely trivial by the relation between the derivatives. The inverse map is simply given by
\begin{equation}	\label{eq.invcoordtransf}
\begin{split}
x^\mu&=y^\mu+\frac{1}{\Lambda^3} \partial^\mu \pi(y)\ ,\\
 \frac{\partial q(x)}{\partial x^\mu} & = \frac{\partial \pi(y)}{\partial y^\mu}\ ,\\
q(x)&=\pi(y)+\frac{1}{2\Lambda^3} (\partial \pi(y))^2 \ .
\end{split}
\end{equation}
Thus the invertibility of the coordinate transformation is also the condition of the invertibility of the map between the two cosets. In the following we are going to study configurations for which the map is invertible.

The transformation (\ref{eq.coordtransf}) can also be thought of as a shortcut for writing a field redefinition with an infinite number of derivatives. If one thinks to leave the coordinates untouched (say that we want to use the $y$ coordinates), the field redefinition takes the form
\begin{equation}	\label{eq.implicitfieldred}
\begin{split}
\pi(y)=q(y+ \Lambda^{-3}\partial \pi(y))-\frac{1}{2\Lambda^3} \left( \partial \pi (y) \right)^2 \ ,
\end{split}
\end{equation}
which can be solved perturbatively. Up to second order in the expansion parameter $\partial^2 q/ \Lambda^{3}$, the field redefinition reads
\begin{equation} \label{eq.fieldred}
\begin{split}
\pi(y)=q(y)+\frac{1}{2\Lambda^3} (\partial q(y))^2+ \frac{1}{2\Lambda^6} \partial^\mu\partial^\nu q(y)\partial_\mu q(y)\partial_\nu q(y)+\dots
\end{split}
\end{equation}
To map one set of Galileon operators into the dual one, one is free to choose one of the two methods in eq.~(\ref{eq.coordtransf}) or eq.~(\ref{eq.fieldred}), but eventually they will give the same result. 

In what follows we will refer to the two non-linear realizations parametrized by the field $\pi$ and $q$ simply as  the $\pi$ and $q$ representations, respectively.

%{}~%%%%%%%%%%%%%%%%%%%%%%%%%%%%%%%%%%%%%
\section{What about Superluminal Propagation?}\label{sec.superluminal}\label{subsec:leadingorder}

In section \ref{sec:TwistAlgebra} we have constructed the $\pi$ and $q$ representations of the coset Gal$(3+1,1)/$ISO$(3,1)$.
At fixed spacetime coordinates, they are related to each other by the field redefinition (\ref{eq.fieldred}) that involves an infinite number of higher derivative terms and an increasing number of 
fields. For any physical process for which the series can be truncated we expect the equivalence of the two representations, since physical observables 
are field redefinition independent. 
However, we will see that there are situations, which are within the regime of validity of the Effective Field Theory (EFT), in which the field redefinition cannot be expanded so that the two representations have a different notion of locality.
A particularly useful and simple example to discuss these issues is the analysis of the propagation of fluctuations in presence of  a non-trivial background.

Our starting point is the free theory in the $\pi$ representation. In \cite{deRham:2013hsa} it was noticed that the field redefinition discussed in the previous section maps a free theory into a linear combination of Galileon operators. The map is the following (see appendix \ref{app:TdR}, eq.~(\ref{DualityMap}))
\begin{equation}	\label{eq.fasmap}
\begin{split}
\mathcal{L}_{G\pi} = \mathcal{L}_{G\pi 2} = -\frac{1}{2}(\partial\pi)^2 \longrightarrow  \ \ \ \mathcal{L}_{Gq} = \mathcal{L}_{Gq2} -\mathcal{L}_{Gq3}+\frac{1}{2}\mathcal{L}_{Gq4} -\frac{1}{6}\mathcal{L}_{Gq5}\,,
\end{split}
\end{equation}
where the ${\cal L}_{Gqi}$ terms are defined in eq.~(\ref{eq.galileons}), with $\pi\rightarrow q$.

Let us analyze the propagation of perturbations $q$ around a given background $\bar q$. We take this to be time-independent and assume that $\partial^2 \bar q / \Lambda^3\ll 1$, working in this section at linear order in this parameter. We are thus interested in terms in the Lagrangian which are quadratic in $q$ and (at most) linear in $\bar q$, therefore we can drop the quartic and quintic Galileons for the rest of this section.
In this regime, the equation of motion for the perturbations is
\begin{equation} \label{eq.motionback}
\begin{split}
\square q -\frac{2}{\Lambda^3}\square \bar q \,\square q+ \frac{2}{\Lambda^3}\partial^\mu\partial^\nu \bar q \,\partial_\mu\partial_\nu q=0 \ .
\end{split}
\end{equation} 
At the order we are working, we can actually drop the second term if we multiply the whole equation by $(1+ 2 \square\bar q/\Lambda^3)$. The wavevector of a perturbation with a wavelength much shorter than the scale of variation of the background satisfies
\begin{equation}
\begin{split}\label{effg}
G^{\mu\nu}k_\mu k_\nu =0\,, \qquad G^{\mu\nu}\equiv \eta^{\mu\nu}+ \frac{2}{\Lambda^3} \partial^\mu\partial^\nu \bar q \,.
\end{split}
\end{equation}
This is equivalent to saying that perturbations are null geodesics of the {\it inverse} metric $G_{\mu\nu}$:\footnote{Notice that eq.~\eqref{eq.motionback} does {\it not} have the same form of the Klein-Gordon equation $\partial_\mu(\sqrt{-G} G^{\mu\nu} \partial_\nu q) =0$. The difference between the two, however, consists of terms where the derivatives hit the background instead of the propagating wave. These are suppressed in the limit in which the wave is very short compared to the scale of variation of $\bar q$, i.e.~in the limit of geometrical optics.}
\be\label{eq:metric}
G_{\mu\nu} \simeq \eta_{\mu\nu} - \frac{2}{\Lambda^3} \partial_\mu\partial_\nu \bar q \;.
\ee 
Null trajectories with respect to this metric will in general be superluminal with respect to the Minkowski lightcone. Indeed, outside a source, where $\nabla^2 \bar q =0$, it is easy to realize that the matrix $\partial_i\partial_j \bar q$ cannot induce subluminal propagation in all directions \cite{Adams:2006sv,Nicolis:2009qm}. This is somewhat disturbing since we are talking about a theory obtained by a field redefinition of a free theory.

Let us look at the geodesic equation. Its time component, since there is no time dependence, simply reduces to $\rmd^2 t/\rmd\lambda^2 =0$, that says we can use $t$ as affine parameter.
We can assume without loss of generality that the propagation, at zeroth order in $\partial^2\bar q/\Lambda^3$, is in one particular direction, say $y$. 
The spatial components of the geodesic equation in terms of the displacement $\delta y^i(t)$ from the unperturbed trajectory, $y=t$, read 
\be
\frac{\rmd^2 \delta y^{i}}{\rmd t^2} = - \Gamma^i_{yy} \frac{\rmd y}{\rmd t}\frac{\rmd y}{\rmd t} =  \frac{1}{\Lambda^3} \partial_{y} \partial_{y} \partial^i \bar q(y = t) \;,
\ee
where we have used $\rmd y/\rmd t =1$ for the unperturbed trajectory and calculated the $\Gamma$'s from the explicit form of the metric in eq.~\eqref{eq:metric}. 
Integrating we get the displacement from the underlying Minkowski lightcone
\begin{equation}\label{deltay}
\begin{split}
\delta y^i = \frac{1}{\Lambda^3} \int \rmd t \int  \rmd t \, \partial_{y} \partial_{y}  \bar q (y = t) =  \frac{1}{\Lambda^3} \partial^i \bar q\Big |_{y_i}^{y_f} \ .
\end{split}
\end{equation}
\begin{figure}[!t]
\begin{center}
		\includegraphics[width=60mm]{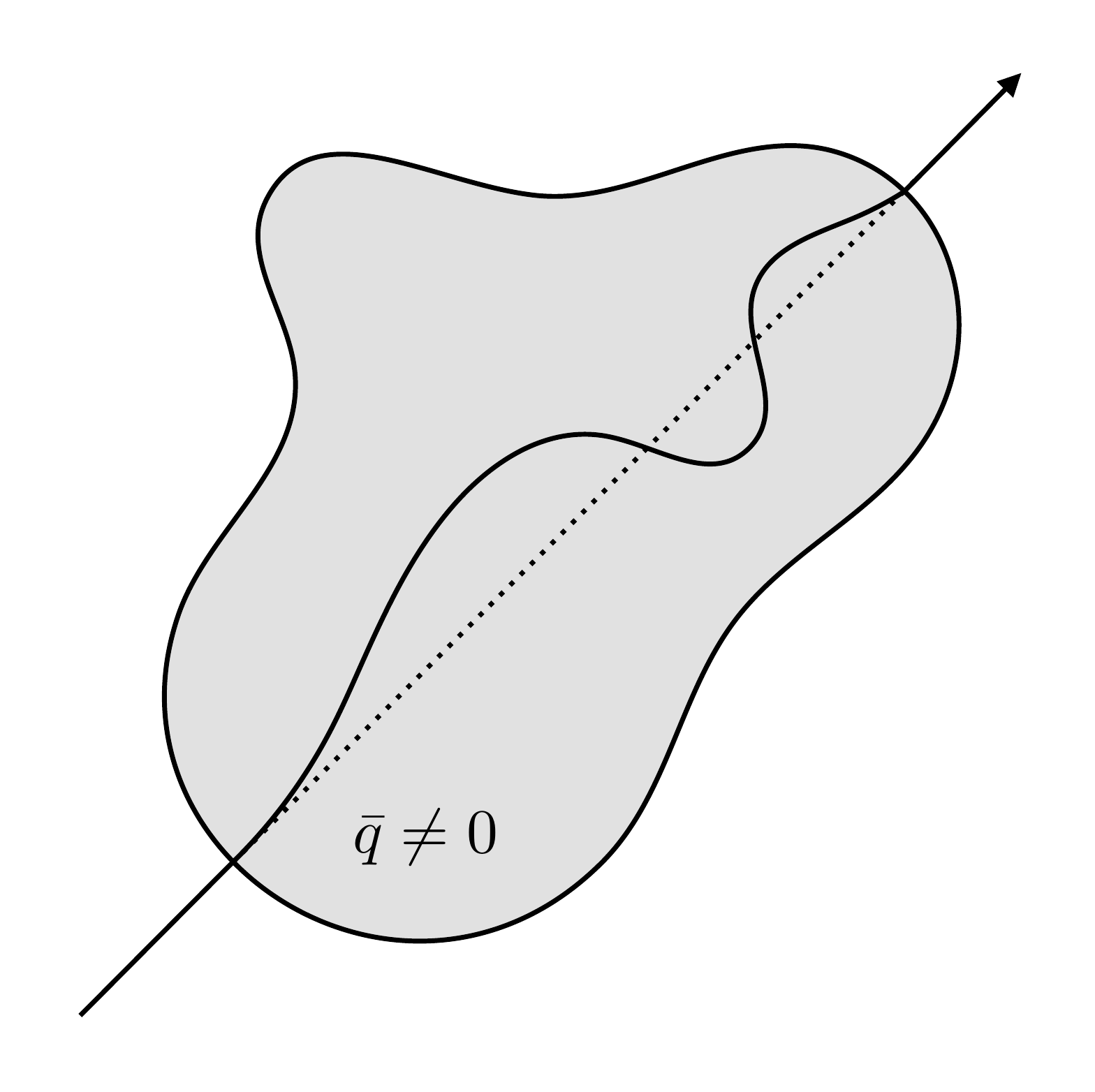}
\end{center}
\caption{\small The deviation from the lightcone averages to zero asymptotically in the theory obtained with a field redefinition of the free theory.}
\label{fig:bkgdev}
\end{figure}
The integrand is a total derivative so that we only get boundary terms. This implies what we should have expected: {\it there is no effect (superluminal or not) if the background $\bar q$ is localized}. There is a deviation from the lightcone as long as we are inside the $\bar q$ background, but the effect averages to zero asymptotically as shown in figure \ref{fig:bkgdev}. The two theories are related by a non-linear field redefinition which changes the interaction terms. Once there are no more interactions with the background (i.e. when the background turns off), the position of a wave-packet must be the same in the two theories, and indeed there is no ``asymptotic superluminality". This is exactly the same argument one uses to show that the S-matrix is the same.  In section \ref{sec.more} we are going to generalize this calculation to higher order in $\partial^2\bar q/\Lambda^3$ and include sources along the path. Furthermore we will show that {\it only for the particular choice of the coefficients} of the eq.~\eqref{eq.fasmap}, which are related to the free theory in the $\pi$ variable, there is a cancellation of any asymptotic effect.

This, however, does not completely clarify the puzzle. What forbids us to measure a ``local superluminality", even in the absence of an asymptotic effect?
In order to measure it, the shift $|\delta y|$ has to exceed at least one wavelength of the perturbation. This means that 
\begin{equation}
\begin{split}\label{eq:measurable}
|\delta y| = \frac{1}{\Lambda^3} |\partial \bar q| \gtrsim \omega^{-1} \,,
\end{split}
\end{equation}
where $\omega$ is the frequency of the perturbation. Since the frequency of the perturbation cannot exceed the cutoff scale,\footnote{In the case of the free theory we are discussing, the scale $\Lambda$ 
does not of course correspond to a cut-off scale, since the theory is UV complete, and one can stick to the weaker condition, eq.~(\ref{eq:measurable}). In the general case, however, $\Lambda$ is the scale above which the non-linear sigma model requires a UV completion.}  this requires
\begin{equation}
\label{eq:EFTval}
\frac{\partial \bar q}{\Lambda^2} \gtrsim 1 \,.
\end{equation}
Notice that eqs.~(\ref{eq:measurable}) and (\ref{eq:EFTval}) do not contradict the working hypothesis $\partial^2\bar q/\Lambda^3 \ll 1$.
This is the usual argument of superluminal propagation \cite{Adams:2006sv,Nicolis:2009qm} in the DGP and Galileon models, which is used to argue against a standard (local and Lorentz invariant) UV completion for this kind of theories. But surely we cannot rule out a free theory!

The key point here is that one has to specify which are the local operators of the theory.  This is usually taken for granted. When we specify a theory, say the free theory, we implicitly assume that the field we use is a local operator and that  it couples locally to other fields. Let us trace back what happens in our case starting from the free field Lagrangian with $\pi$ locally coupled to an external source $J$:
\begin{equation}
\begin{split}
\mathcal{L}_{G\pi} =\mathcal{L}_{G\pi 2}+\mathcal{L}_{J} =-\frac{1}{2}(\partial\pi)^2 + \pi(y) J(y) \ .
\end{split}
\end{equation}
In discussing superluminality, it is perhaps more intuitive to keep the same coordinates and only transform the field, as in eq.~\eqref{eq.fieldred}. 
Since eq.~(\ref{eq.fieldred}) contains an infinite number of terms, in general the mapped Lagrangian will also contain an infinite number of operators. 
Only for Galileon operators, quite remarkably, we get a Lagrangian with a finite number of terms, as in eq.~(\ref{eq.fasmap}), all the remaining terms being total derivatives. 
In particular, infinitely many operators arise when we map $\mathcal{L}_{J}$  in the $q$-representation: 
\begin{equation}\label{LJsource}
\begin{split}
\mathcal{L}_{J}= \pi(q(y)) J(y)=\left( q(y)+\frac{1}{2\Lambda^3} (\partial q(y))^2+ \frac{1}{2\Lambda^6} \partial^\mu\partial^\nu q(y)\partial_\mu q(y)\partial_\nu q(y)+\dots\right)J(y) \ .
\end{split}
\end{equation}
From an EFT point of view,  interpreting the scale $\Lambda$ as a cut-off,  one would naively be justified in truncating the series and dismiss higher derivative operators. 
The series can indeed be truncated when considering a perturbative scattering amplitude involving a given, finite, number of fields in a trivial background, since
the higher order terms in the series involve an increasing number of fields.

However, the series {\it cannot} be truncated in discussing whether $q$ propagates superluminally around a non-trivial background.
Indeed, the Taylor series (\ref{eq.fieldred}) contains, among others, terms of the form 
\begin{equation}
\begin{split}\label{eq:terms2keep}
\partial^n q\cdot \left(\frac{\partial \bar q}{\Lambda^3}\right)^n \simeq \left(\frac{\omega \partial\bar q}{\Lambda^3}\right)^n q  \,.
\end{split}
\end{equation}
Given the criterium \eqref{eq:measurable} for a measurable superluminality, 
we must keep all the terms in eq.~(\ref{eq:terms2keep}).
Interestingly enough, their contribution can be resummed. This is easily seen by expanding  eq.~(\ref{eq.implicitfieldred}):
\be
\pi(y) = q(y)+ \sum_{n=1}^\infty \frac{1}{n!}\frac{(\partial \pi(y))^n (\partial^n q(y))}{\Lambda^{3n}}-\frac{1}{2\Lambda^3}(\partial \pi(y))^2 \,,
\label{piqyExp}
\ee
where for simplicity we have omitted the obvious Lorentz tensor structure that appears in the derivative terms. Let us focus on the terms linear in the fluctuations 
and with one derivative acting on $\bar q$. For $n\geq 2$ in the above sum the fluctuation must be provided by the term $\partial^n q$, while $(\partial \pi)^n = (\partial\bar q)^n$  
gives the background contribution. The resulting series is easily resummed, leading to a non-local coupling with the source
\begin{equation}	\label{eq.nonloc1}
\begin{split}
\mathcal{L}_{J}\supset q\left(y^\mu+ \frac{1}{\Lambda^{3}}\partial^\mu \bar q(y)\right) J(y) \ .
\end{split}
\end{equation}
Notice how the last term in eq.~(\ref{piqyExp}) is crucial to be able to recast the coupling in the form (\ref{eq.nonloc1}).
Now we see that there is no paradox. Even though the theory ${\cal L}_{Gq}$ allows for superluminal propagation, it is equivalent to the free theory we started with {\it only if} we use non-local couplings of the form \eqref{eq.nonloc1}. It is straightforward to see that these non-local couplings exactly cancel the change in signal propagation given by eq.~(\ref{deltay}). If, on the other hand, we use local couplings of sources with ${\cal L}_{Gq}$, we have superluminality, but {\it this theory is different from the free one we started with.}

\begin{figure}[!t]
\begin{center}
		\includegraphics[width=100mm]{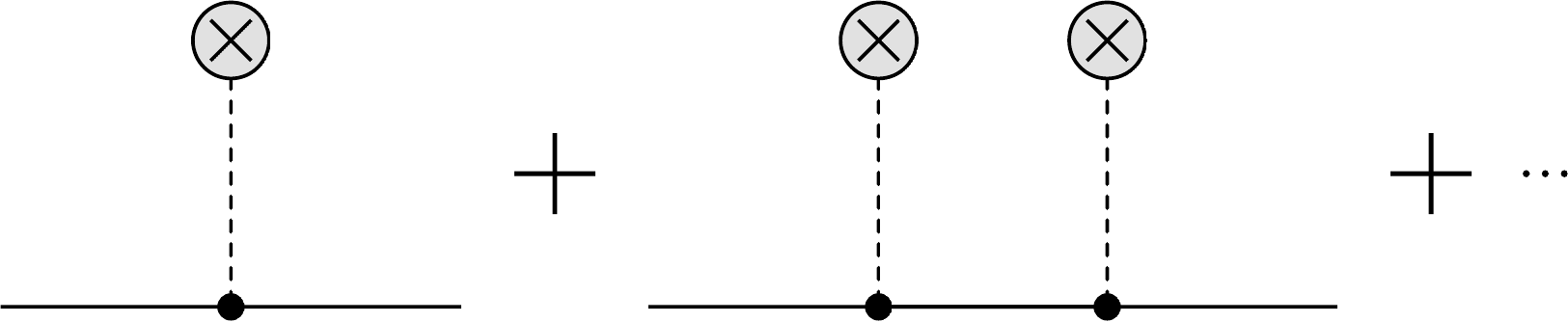}
\end{center}
\caption{\small Superluminality, even at leading order in $\partial^2 \bar q/\Lambda^3$, involves the sum of infinitely many diagrams.}
\label{fig:propagator}
\end{figure}
It can be useful to see how the deviation (\ref{deltay}) from a luminal behaviour would arise in a diagrammatic approach. 
The background field $\bar q$ acts as a disturbance to the propagation of the 
fluctuation of $q$ itself, changing its dispersion relations.  At leading order in $\partial^2 \bar q / \Lambda^3$, the leading interaction arises from ${\cal L}_{Gq3}$.
Integrating by parts, we can write the interactions as
\be
{\cal L}_{Gq} \supset \partial^\mu q \partial^\nu q \partial_\mu\partial_\nu \bar q - (\partial q)^2 \Box \bar q\,.
\label{IntLag}
\ee
It is important to stress that when studying the propagation of $q$ in the background $\bar q$ we are not considering a single vertex of eq.~\eqref{IntLag}, but we are resumming an infinite set of diagrams, as we show in figure \ref{fig:propagator}. This is done automatically in keeping $\bar q$ in the modified kinetic term for the fluctuations. 
The point is that when we write an expression like 
\begin{equation}	\label{eq.wave}
\begin{split}
q(x)\propto e^{- i k(1+\delta c) t +i \vec k \cdot \vec x} \ ,
\end{split}
\end{equation}
with $\delta c$ the variation of the speed of propagation with respect to the speed of light, and we keep $\delta c$ at the exponent without linearizing, we are making a non-perturbative statement with an infinite number of $\bar q$. The criterium for measurable superluminality is saying that the effect of $\delta c$ at the exponent is large so that the exponential cannot be expanded. That is why superluminality is a statement about an infinite number of diagrams.

In section \ref{subsec:allorders} we extend our result at all orders in $\partial^2\bar q/\Lambda^3$ for particular backgrounds in which the matrix $\partial_i\partial_j \bar q$ is constant, so that the dynamics of perturbations is effectively translationally invariant.

\section{\label{sec.more}More on Asymptotic Effects and Locality}
We give here several additional examples which involve asymptotic effects and locality. This section can be skipped without loss of continuity.
\subsection{\label{sub.secondorder}Asymptotic Effects at Second Order in the Background}
In this subsection we generalize the propagation of signals at second order in $\partial^2\bar q/\Lambda^3$ and show the absence of any asymptotic effect in the theory \eqref{eq.fasmap}, obtained by mapping the free theory. At this order also the quartic Galileon is relevant, so that we can check that its precise coefficient in eq.~\eqref{eq.fasmap} is crucial for the cancellation of asymptotic effects. The equation of motion can be written as
\begin{equation} \label{eq.motionback2}
\begin{split}
\square q + \frac{2}{\Lambda^3}\partial^\mu\partial^\nu \bar q \,\partial_\mu\partial_\nu q + \frac{3}{\Lambda^6} \partial^\mu\partial^\rho\bar q \partial_\rho\partial^\nu\bar q \partial_\nu\partial_\mu q=0 \ .
\end{split}
\end{equation} 
To get to this form we used the equation of motion of the background $\bar q$, assuming there are no sources in the region we consider. At lowest order the background thus satisfies $\square\bar q =0$. This corresponds to the propagation on an effective metric
\be
G^{\mu\nu} = \eta^{\mu\nu} + \frac{2}{\Lambda^3}\partial^\mu\partial^\nu \bar q+ \frac{3}{\Lambda^6} \partial^\mu\partial_\rho\bar q \partial^\rho\partial^\nu \bar q \;,
\ee
and thus
\be
G_{\mu\nu} \simeq \eta_{\mu\nu} - \frac{2}{\Lambda^3}\partial_\mu\partial_\nu \bar q+ \frac{1}{\Lambda^6} \partial_\mu\partial_\rho\bar q \partial^\rho\partial_\nu\bar q \;.
\ee
For a time-independent background we can still use $t$ as affine parameter and the second order contributions to the geodesic equation are
\be
\frac{\rmd^2 \delta y^{(2)\;i}}{\rmd t^2} + \Gamma^{(2)\;i}_{yy} + \partial_j \Gamma^{(1)\;i}_{yy} \delta y^{(1)\;j} + 2 \Gamma^{(1)\;i}_{jy} \frac{\rmd \delta y^{(1)\;j}}{\rmd t} = 0 \,,
\ee
where the number in parentheses indicate the order in $\partial^2\bar q/\Lambda^3$. The third term in the equation takes into account that the Christoffel symbols are evaluated on the first order geodesic. Once this is taken into account all terms in the equation are evaluated on the unperturbed trajectory in the $y$ direction: $y = t$.  Evaluating all the $\Gamma$'s and using the previous result on $\delta y^{(1)\;i}$, eq.~(\ref{deltay}), it is straightforward to get 
\be
\frac{\rmd^2 \delta y^{(2)\;i}}{\rmd t^2} = \partial_y\partial_y \left(\partial^i\partial_j\bar q \partial^j \bar q\right) \;.
\ee
As at first order, the integral is a total derivative and it vanishes exactly if we consider localized backgrounds that vanish at infinity.

%%%%%%%%%%%%%%%%%%%%%%%%%%
\subsection{\label{sub.cylinder}Asymptotic Effects for a Background with Cylindrical Symmetry}

Another interesting example is the one of a background solution $\bar q$ with cylindrical symmetry. Even though the propagation of fluctuations on the axis of the symmetry reduces to a one dimensional problem (due to symmetry the geodesic cannot deviate), the background $\bar q$ depends on all three spatial coordinates so that all the Galileons $\mathcal{L}_{Gi}$ are relevant. This example will show that only the particular combination given by eq.~(\ref{eq.fasmap}) gives no asymptotic effect. We will work up to cubic order in $\partial^2\bar q/\Lambda^3$, which is the first non-trivial order that includes $\mathcal{L}_{G5}$.

With this setup in mind, the equation of motion can be written as
\begin{equation} \label{eq.eomcilinder}
\begin{split}
\square q + \partial_\mu\partial_\nu q\left(\frac{2}{\Lambda^3}\partial^\mu\partial^\nu \bar q \, + \frac{3}{\Lambda^6}  \partial^\mu\partial_\rho\bar q  \,\partial^\rho\partial^\nu\bar q + \frac{4}{\Lambda^9}  \partial^\mu\partial^\rho\bar q \, \partial_\rho\partial_\sigma\bar q  \,\partial^\sigma\partial^\nu\bar q\right)=0 \ ,
\end{split}
\end{equation} 
where we assumed that there are no sources along the axis of symmetry so that the background satisfies the equation of motion given by the Lagrangian (\ref{eq.fasmap}). We assume that the propagation of perturbations happens on the symmetry axis (say, the z-axis) of the background $\bar q$. Consider then a wave with energy $\omega$ and momentum $k_z$. Its dispersion relation reads
\begin{equation}
\begin{split}
\omega^2=k_z^2\left(1+\frac{2}{\Lambda^3}\partial^z\partial^z \bar q \, + \frac{3}{\Lambda^6}  \partial^z\partial_\rho\bar q  \,\partial^\rho\partial^z\bar q + \frac{4}{\Lambda^9}  \partial^z\partial^\rho\bar q \, \partial_\rho\partial_\sigma\bar q  \,\partial^\sigma\partial^z\bar q\right)\, .
\end{split}
\end{equation}
Furthermore, using cylindrical coordinates it is easy to see that all contractions vanish but the ones which involve the $z$-coordinate. This is because both $\partial_r \bar q=0$ and $\partial_\theta \bar q=0$ on the $z$-axis, thus
\begin{equation}
\begin{split}
\omega^2=k_z^2\left(1+\frac{2}{\Lambda^3}\partial_z^2 \bar q \, + \frac{3}{\Lambda^6} ( \partial_z^2\bar q)^2 + \frac{4}{\Lambda^9}  (\partial_z^2\bar q)^3\right) \, ,
\end{split}
\end{equation}
and the speed of propagation along the $z$-axis is
\begin{equation}
\begin{split}
v_z(z)=\frac{d\omega}{d k_z}=\left(1+\frac{2}{\Lambda^3}\partial_z^2 \bar q \, + \frac{3}{\Lambda^6} ( \partial_z^2\bar q)^2 + \frac{4}{\Lambda^9}  (\partial_z^2\bar q)^3\right)^{1/2} \, .
\end{split}
\end{equation}
Since the propagation is in $1+1$ dimensions, and the velocity depends only on $z$, it is easier to integrate in the spatial coordinate instead of time. This leads to
\begin{equation}
\begin{split}
\Delta t=\int_{z_i}^{z_f} dz \,\frac{1}{v_z(z)}\simeq\Delta z-\int_{z_i}^{z_f} dz \,\frac{1}{\Lambda^3}\partial^2_z \bar q  =\Delta z-\frac{1}{\Lambda^3}\partial_z \bar q\Big|_{z_i}^{z_f}+\mathcal{O}(\bar q^4)\, .
\end{split}
\end{equation}
The precise coefficients of the quartic and quintic operators in the Lagrangian (\ref{eq.fasmap}) is such that only the first order term remains. Moreover, if the background is localised (asymptotically trivial), even the linear contribution goes to zero and $\Delta t=\Delta z$. Of course, this comes  with no surprise: this choice of couplings is what makes the theory free. 

%%%%%%%%%%%%%%%%%%%%%%%%%
\subsection{\label{sub.sources}Asymptotic Effects Including Sources}
In this subsection we show that there is no asymptotic superluminality in a general $1+1$ dimensional problem with time independent backgrounds. In this setup we are forced to drop the assumption made so far that there are no sources along the path of propagation. We will work up to second order in $\partial^2 \bar q/ \Lambda^3$. In $1+1$ dimensions 
the Galileon operators ${\cal L}_{Gq4}$ and ${\cal L}_{Gq5}$ cancel exactly and we are left with $\mathcal{L}_{Gq i}$, with $i=1,2,3$. In $1+1$ dimensions we cannot avoid to add some tadpole term $\pi(y) \bar J(y)$ for generating the background $\bar \pi$, which will satisfy the equation of motion
\begin{equation}\label{eq.bkg1dpi}
\begin{split}
\square \bar \pi =-\bar J(y)\, .
\end{split}
\end{equation}
Mapping the kinetic term and the tadpole in the $\pi$-representation by means of eq.~(\ref{eq.fieldred}) in the $q$-representation, we obtain 
\begin{equation}	\label{eq.fasmapwithJ}
\begin{split}
\mathcal{L}_{G\pi} = -\frac{1}{2}(\partial\pi)^2 +\pi \bar J(y)\longrightarrow  \  \mathcal{L}_{Gq} =\mathcal{L}_{Gq2} -\mathcal{L}_{Gq3}+ \left( q+\frac{1}{2\Lambda^3} (\partial q)^2+ \frac{1}{2\Lambda^6} \partial^\mu\partial^\nu q\partial_\mu q\partial_\nu q+\dots\right)\bar J(y)\, .
\end{split}
\end{equation}
Once the equation of motion is expanded around the background solution 
\begin{equation}\label{eq.bkg1dq}
\begin{split}
\square \bar q =-\bar J(y)+\mathcal{O}(\bar q^2)\, ,
\end{split}
\end{equation}
the dispersion relation for fluctuations will involve also derivatives acting on $\bar J(y)$, but in general these terms can be neglected since the background is assumed to be slowly varying with respect to the fluctuations. The dispersion relation for perturbations thus reads
\begin{equation}
\begin{split}
\omega^2=k^2\left( 1+\frac{2}{\Lambda^3}\partial^2 \bar q+\frac{1}{\Lambda^3}\partial^2\bar q \bar J+\frac{4}{\Lambda^6}(\partial^2\bar q)^2 \right)+\mathcal{O}(\bar q^3)\,,
\end{split}
\end{equation}
and using eq.~(\ref{eq.bkg1dq}) the speed of propagation is
\begin{equation}
\begin{split}
v(y)=\frac{d\omega}{d k}=\left( 1+\frac{2}{\Lambda^3}\partial^2 \bar q+\frac{3}{\Lambda^6}(\partial^2\bar q)^2 \right)^{1/2}\simeq1+\frac{1}{\Lambda^3}\partial^2 \bar q+\frac{1}{\Lambda^6}(\partial^2\bar q)^2+\mathcal{O}(\bar q^3)\,.
\end{split}
\end{equation}
This leads to
\begin{equation}
\begin{split}
\Delta t=\int_{y_i}^{y_f} dy \,\frac{1}{v(y)}\simeq\Delta y-\int_{y_i}^{y_f} dy \,\frac{1}{\Lambda^3}\partial^2 \bar q  =\Delta y-\frac{1}{\Lambda^3}\partial \bar q\Big|_{y_i}^{y_f}+\mathcal{O}(\bar q^3)\, .
\end{split}
\end{equation}
If the background vanishes asymptotically this contribution goes to zero and $\Delta t=\Delta y$, as expected. 

%%%%%%%%%%%%%%%%%%%%%%%%%
\subsection{\label{sub.DGP}Asymptotic Superluminality in the DGP model}
As it was already noticed in \cite{Adams:2006sv, Nicolis:2004qq} the DGP model allows for superluminal propagation around non-trivial backgrounds. In this section we show that the superluminality persists even asymptotically; in particular we look at spherically symmetric backgrounds which vanish at infinity.

In the decoupling limit the Lagrangian for the Goldstone of the DGP model reads \cite{Dvali:2000hr, Nicolis:2004qq}
\begin{equation}\label{eq.DGP}
\begin{split}
\mathcal{L}_{DGP}=6 \mathcal{L}_{Gq2}+2\mathcal{L}_{Gq3}+\frac{1}{2M_4}qT\,,
\end{split}
\end{equation}
where $T$ is the trace of the stress energy tensor and $M_4$ is the four dimensional Planck scale. From this, the equation of motion is
\begin{equation}\label{eq.dgpstar}
\begin{split}
3 \square q  +\frac{1}{\Lambda^3}(\square q)^2- \frac{1}{\Lambda^3}\partial^\mu\partial^\nu  q \,\partial_\mu\partial_\nu q=-\frac{1}{4M_4}T \, .
\end{split}
\end{equation} 
Consider a  spherically symmetric and homogeneous distribution of matter inside a radius $R_\star$ (much larger than the cut-off length $\Lambda^{-1}$), with total mass $M_\star$.
The exact  solution of eq.~(\ref{eq.dgpstar}) for a spherically symmetric source can easily be found using Gauss theorem and it reads\footnote{Here we have already chosen the stable solution. See \cite{Nicolis:2004qq} for details.}
\begin{equation}
\begin{split}
\partial_r \bar q(r)=E(r) =
\begin{cases}
\frac{3\Lambda^3}{4}r\left( \sqrt{1+\frac{1}{2\pi}\frac{R_V^3}{R_\star^3}} -1\right), & \text{if }r\leq R_\star \\
\frac{3\Lambda^3}{4}r\left( \sqrt{1+\frac{1}{2\pi}\frac{R_V^3}{r^3}} -1\right), & \text{if }r> R_\star
\end{cases}
\end{split}
\end{equation}
where $R_V=\Lambda^{-1}(M_\star/M_4)^{1/3}$ is the Vainshtein radius for the source.
Once the action given by eq.~(\ref{eq.DGP}) is expanded around the background solution, the perturbations of the field $q$ will travel on null geodesics of the effective metric \cite{Nicolis:2004qq}
\begin{equation}
\begin{split}
G^{\mu\nu}=\left(3+\frac{2}{\Lambda^3}\frac{\partial_r(r^2 E)}{r^2} \right)g^{\mu\nu}- \delta^\mu_r\delta^\nu_r \frac{2}{\Lambda^3}\partial_rE- \frac{\delta^\mu_\theta \delta^\nu_\theta}{r^2\,\text{sin}^2\phi} \frac{2}{\Lambda^3}\frac{E}{r}
- \frac{\delta^\mu_\phi \delta^\nu_\phi}{r^2} \frac{2}{\Lambda^3}\frac{E}{r}\end{split}\,,
\end{equation}
where $g^{\mu\nu}$ is the Minkowski metric in spherical coordinates. Here we consider a radial null geodesic passing through the center of the compact object. It is given by the condition
\begin{equation}
\begin{split}
ds^2=-\left(3+\frac{2}{\Lambda^3}\frac{\partial_r(r^2 E)}{r^2} \right)^{-1}dt^2+\left(3+\frac{4}{\Lambda^3}\frac{ E}{r} \right)^{-1}dr^2=0\,,
\end{split}
\end{equation}
which defines the radial velocity
\begin{equation}
\begin{split}
v_r=\dot r=\left( 1-\frac{\frac{2}{\Lambda^3}\partial_rE}{3+\frac{2}{\Lambda^3}\frac{\partial_r(r^2 E)}{r^2}}\right)^{1/2} \,.
\end{split}
\end{equation}
The novelty with respect to \cite{Nicolis:2004qq} is that in the inner region $r<R_\star$, the constant density produces a background $\bar q \propto r^2$ which makes the fluctuation to propagate with a constant and subluminal speed given by
\begin{equation}\label{eq.radvel}
\begin{split}
v_r=\sqrt{1-\frac{c_1}{3c_1+2}}\,, \qquad c_1=\sqrt{1+\frac{R_V^3}{2\pi R_\star^3}} -1\,.
\end{split}
\end{equation}
\begin{figure}[!t]
\begin{center}
		\includegraphics[width=0.6\textwidth]{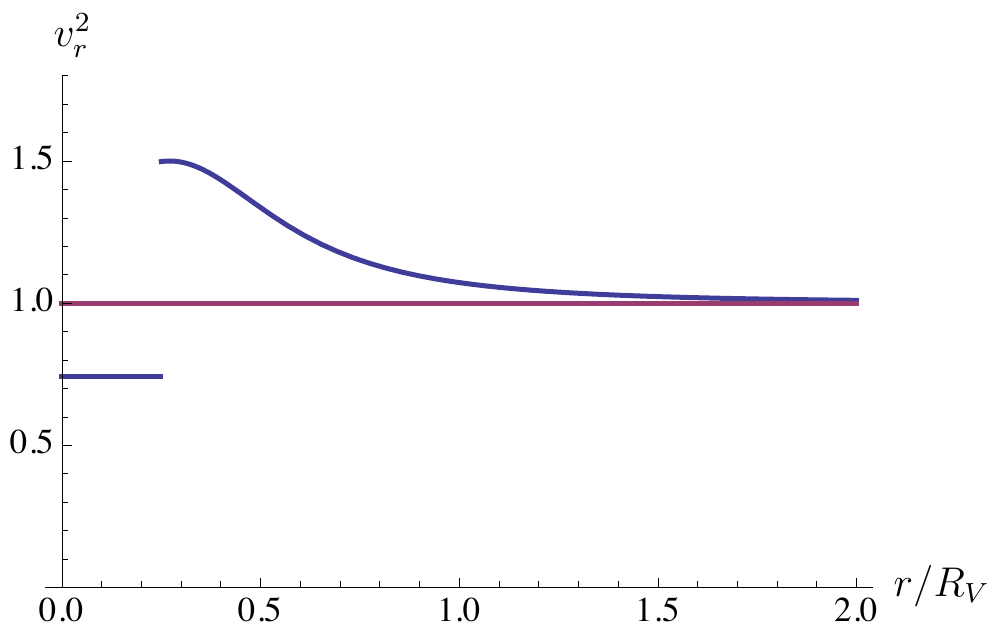}
\end{center}
\caption{\small Plot of the speed of propagation as a function of $r$ for $R_\star=0.25 R_V$. The discontinuity in the velocity is produced by the abrupt change in the density field and can be removed by smoothing the density $\rho_\star$.}
\label{fig:DGPspeed}
\label{fig.speed}
\end{figure}

As can be seen from figure~\ref{fig.speed}, in the inner region the speed is smaller than $1$, while outside the source it is greater than 1 and it approaches 1 asymptotically from above. Thus to have asymptotic superluminality  it is sufficient to have a source with a radius $R_\star$ much smaller than its Vainshtein radius $R_V$ (notice that the speed (\ref{eq.radvel}) remains finite even when the density diverges), 
in this case
the outer region dominates  and there is superluminal propagation. In the limit $R_\star \ll R_V$ the inner region can be neglected and the advance in the propagation is given by
\begin{equation}
\begin{split}
\delta t=2 \int_0^\infty \left(v_r^{-1}-1\right) \,dr\simeq -0.3 \,R_V\,.
\end{split}
\end{equation}
To avoid the discontinuity in the velocity produced by the abrupt change in the density field, one can imagine to smooth the density across a thin layer around $R_\star$, but this does not change the picture qualitatively.

%%%%%%%%%%%%%%%%%%%%
\subsection{Non-Locality for a Translationally Invariant Background}
\label{subsec:allorders}

In this subsection we come back to the cancellation between superluminality and non-local couplings with sources. In general it is technically difficult to go beyond the first order calculation of section \ref{sec.superluminal}, but a full non-linear result can be achieved
for backgrounds that are effectively translationally invariant.
As can be seen from eqs.~(\ref{eq.motionback}) and (\ref{IntLag}) these are the ones for which $\partial^2 \bar q$ is constant.
For simplicity, consider a background in the $q$-representation that depends on a single spatial coordinate only,\footnote{To avoid clutter in the notation, we denote with the same symbol $x$ the spatial coordinate  the background depends on,  and the four-vector $x^\mu$. The distinction should be clear from the context. Similarly for the variable $y$ in the $\pi$-representation.}
\begin{equation}	\label{eq.y2bk}
\begin{split}
\bar q(x) = \frac{1}{2} \epsilon \Lambda^3 x^2 \,.
\end{split}
\end{equation}
Using the map (\ref{eq.invcoordtransf}),  eq.~(\ref{eq.y2bk}) is mapped in the $\pi$-representation to the background 
\be
\bar \pi(y) = \frac 12 \kappa \Lambda^3  y^2 \,,
\label{piBack}
\ee
where
\be
\kappa =  \frac{\epsilon}{1-\epsilon} \,.
\ee
The configuration (\ref{piBack}) is obtained in the free theory by adding the tadpole
\begin{equation}\label{eq.tadbac}
\begin{split}
\mathcal{L}_{tad}= -\kappa  \Lambda^4 {\mathcal L}_{G\pi 1} = - \kappa \Lambda^3 \pi \,.
\end{split}
\end{equation}
The total Lagrangian that we are considering is then 
\begin{equation}
\begin{split}
\mathcal{L}_{G\pi} =\mathcal{L}_{G\pi 2}+\mathcal{L}_{tad}+\mathcal{L}_{J} =-\frac{1}{2}(\partial\pi)^2  -\kappa \Lambda^3 \pi +\pi(y) J(y) \ .
\end{split}
\end{equation}
The tadpole (\ref{eq.tadbac}) corrects the coefficients of the Lagrangian in the $q$-representation by terms proportional to the background, and thus there was no need to consider it in the leading order calculation presented in section \ref{subsec:leadingorder}. Under the map (\ref{DualityMap}), we get
\begin{equation}
\begin{split}
\mathcal{L}_{G\pi} \longrightarrow \mathcal{L}_{Gq} = -\kappa \mathcal{L}_{Gq1}+ \left(1+\kappa \right)\mathcal{L}_{Gq2} -\left(1+\frac{\kappa}{2}\right)\mathcal{L}_{Gq3}+\frac{1}{2}\left( 1+\frac{\kappa }{3} \right)\mathcal{L}_{Gq4} -\frac{1}{6}\left( 1+\frac{\kappa}{4} \right)\mathcal{L}_{Gq5}+\mathcal{L}_{J} \ .
\end{split}
\label{Lpiinq}
\end{equation}
The equation of motion for the fluctuations is
\begin{equation}
\begin{split}
(1+\kappa) \square q -2\left(1+\frac{\kappa}{2}\right)\epsilon \,\square q+ 2\left(1+\frac{\kappa }{2}\right)\epsilon \delta^\mu_x\delta_x^\nu \,\partial_\mu\partial_\nu q=0 \,.
\end{split}
\end{equation} 
The speed of propagation of the perturbations can be directly read from the equation of motion to be
\begin{equation}
\begin{split}
v=\frac{1}{1-\epsilon} \ ,
\end{split}
\end{equation}
which for $\epsilon>0$ leads to a ``would-be" superluminal gain
\begin{equation}
\begin{split}
\delta y =\frac{y}{1-\epsilon} \ .
\end{split}
\end{equation} 

Let us now study the non-local coupling with the source. Using the map  (\ref{eq.invcoordtransf}), we have
\be
\int d^4y \, \pi(y) J(y) = \int d^4x\, {\rm Jac}(q(x))\, \Big(q(x) -\frac 1{2\Lambda^3} (\partial q)^2 \Big) J\Big(x-\frac{\partial q}{\Lambda^3}\Big)\,,
\label{SourceJ}
\ee
where
\be
{\rm Jac}(q(x))=\left|{\rm Det}\frac{\partial y^\mu}{\partial x^\nu}\right| = \Big|{\rm Det}\ \Big(\delta^\mu_\nu -\frac{1}{\Lambda^3}\partial^\mu\partial_\nu q\Big)\Big|  = 1-\epsilon +\ldots 
\ee
is the Jacobian factor  associated to the coordinate change (\ref{eq.coordtransf}) and $\ldots$ are terms involving fluctuations of $q$.
Keeping only linear couplings of the fluctuations with the source gives simply
\be \label{source}
\int d^4y \, \pi(y) J(y) =  \int d^4x (1-\epsilon) \, q(x)  J(x(1-\epsilon))\,.
\ee
Possible additional higher derivative terms linear in the fluctuations, present in the  Lagrangian (\ref{SourceJ}), exactly cancels out. 
By rescaling the $x$ coordinate 
\be
x \rightarrow (1+\kappa) x = \frac{1}{1-\epsilon} x
\ee
the Lagrangian (\ref{Lpiinq}) becomes canonically normalized and the source term (\ref{source}) becomes 
\be \label{sourceCan}
 \int d^4x\, q\Big(t,\frac{x}{ 1-\epsilon},y,z\Big)  J(t,x,y,z) + \ldots
\ee
Thus, the field in the point 
$
\left( t, x/(1-\epsilon),0,0 \right)
$,
which  corresponds to a superluminal geodesic, is coupled to a receiver on the lightcone at $x^\mu=(t,x,0,0)$. 
Hence, the non-local couplings induced by the field redefinition cancel exactly the superluminal propagation: as in the free theory, sources exchange signals when their separation is null.

\section{Superluminality and Locality for the Conformal Coset}
\label{sec:conformal}

Let us briefly recall the findings of \cite{Bellucci:2002ji,Creminelli:2013fxa} about the non-linear realizations of the conformal group. 
Using the coset construction, two different non-linear realizations of SO(4,2)/ISO(3,1) can be obtained, denoted DBI and Weyl representation \cite{Creminelli:2013fxa}.
Their contraction leads respectively to the $q$ and $\pi$ representations of Gal($3+1$,1)/ISO(3,1) discussed so far.
The DBI and Weyl representations are related by the following relations  \cite{Bellucci:2002ji}:
\be
y^\mu = x^\mu+ L e^{q(x)/L} \lambda^\mu(x)\,, \ \ \ \ \pi(y) = \frac{q(x)}L + \log(1+\lambda^2(x))\,, \ \ \  \Omega_\mu(y) = \frac 1L \lambda_\mu(x) \,,
\label{TransfEqs}
\ee
and their inverses
\be
x^\mu = y^\mu- L^2 e^{\pi(y)} (1+L^2 \Omega^2(y))\Omega^\mu(y), \ \ \ \frac{q(x)}L = \pi(y) -  \log(1+L^2 \Omega^2(y)), \ \  \lambda_\mu(x)= L \Omega_\mu(y)  \,,
\label{TransfEqsInverse}
\ee
where
\be
\lambda_\mu(x) = \frac{\partial_\mu q(x) \, e^{q(x)/L}}{1+\sqrt{1+e^{2q(x)/L} (\partial q(x))^2}}\,, \ \ \ \ \Omega_\mu(y) = \frac12 e^\pi \partial_\mu\pi(y) \,.
\ee
In \cite{Creminelli:2013fxa} it was observed that the dilaton fluctuations in the DBI representation travel subluminally when 
one expands the leading Nambu-Goto action around a background $\bar q$ of the form
\begin{equation}\label{Cmu}
\begin{split}
L\, \partial_\mu e^{\bar q(x)/L} = C_\mu\,.
\end{split}
\end{equation}
On the other hand, when mapped by means of eq.(\ref{TransfEqsInverse})  in the Weyl representation  the dilaton fluctuations
travel at the speed of light.  Not suprisingly, as we have seen  before for the $q$ and $\pi$ representations of the coset Gal(3+1,1)/ISO(3,1), the resolution of this apparent discrepancy  is due to the different notion of locality in the DBI and Weyl representations.

Let us focus on the background (\ref{Cmu}) where $C_i$ vanish and the background depends on time only. This is the case relevant for the Genesis scenarios \cite{Creminelli:2010ba,Hinterbichler:2012yn}, where we have
\be
\label{eq:SO41back}
L e^{\bar \pi(y)} =-\alpha_\pi y^0\,, \qquad L e^{\bar q(x)/L}=  -\alpha_q x^0\, ,
\ee
with
\be
 \alpha_q = \alpha_\pi \left(1+\frac{ \alpha_\pi^2}4\right)^{-1} \,, \ \ \  x^0 = y^0 \cdot \frac{1+\frac{\alpha_\pi^2}{4 }}{1-\frac{ \alpha_\pi^2}{4 }} \;.
\ee
The simplest action that produces the background (\ref{eq:SO41back}) in the AdS representation contains only the first two conformal DBI Galileons:
\begin{equation}\label{LqNG}
\begin{split}
\mathcal{L}_{q}=-(1-\alpha_q^2)^{-1/2}\mathcal{L}_{q1}+\mathcal{L}_{q2}=-\frac{1}{L^4} e^{-4q/L} \left( \sqrt{1+e^{2q/L}(\partial q)^2} - 
\frac{1}{\sqrt{1-\alpha_q^2}} \right) \,.
\end{split}
\end{equation}
As was already observed in \cite{Hinterbichler:2012yn,Hinterbichler:2012fr}, the quadratic Lagrangian for the fluctuations of the field $\chi=L^{-1}e^{-q/L}$ on the background $\bar{\chi} = -1/(\alpha_q x_0)$ in eq.~(\ref{eq:SO41back}) reads
\begin{equation}
\label{actionchi}
\begin{split}
S_{\chi}^{(2)} = \frac{1}{2}\gamma^3\int\!d^4x \Big(\dot \chi^2 -\frac{1}{\gamma^2} (\partial_i \chi)^2 +\frac{6}{x_0^2}\chi^2\Big) \,,
\end{split}
\end{equation}
where 
\be
\gamma^2 = \frac1{1-\alpha_q^2} \;, 
\ee
and the fluctuations travel strictly subluminally with speed of sound\footnote{The mass term in (\ref{actionchi}) must be present for symmetry reasons: it cancels the variation of the kinetic term under time translation $\delta_{P_0} \chi = \bar{\chi}/t$ that are non-linearly realized because the background is time dependent. However in the following, when we discuss the propagation, it can be neglected since we consider perturbations with typical frequencies much higher than the inverse time scales on which the background varies.}
\begin{equation}
\begin{split}
c_s=\gamma^{-1}=\sqrt{1-\alpha_q^2}\,.
\end{split}
\end{equation}
Using the map (\ref{confmap}) obtained in \cite{Creminelli:2013fxa}, we get the Lagrangian in the Weyl representation
\begin{equation}
\begin{split}
\mathcal{L}_{q}\longrightarrow \mathcal{L}_{\pi}=\left(1-\gamma \right) \mathcal{L}_{\pi1}+\frac{1}{2} \left( 2\gamma- 1\right)\mathcal{L}_{\pi2}-\frac{1}{4} \gamma \mathcal{L}_{\pi3}+\frac{1}{48}\left(2 \gamma+ 1 \right)\mathcal{L}_{\pi4}-\frac{1}{192}\left( 1+\gamma \right)\mathcal{L}_{\pi5} \,.
\end{split}
\end{equation}
The quadratic action for the fluctuation of the field $\phi=L^{-1}e^{-\pi}$ reads\footnote{Here $N(\alpha_q)$ is just some overall normalisation which does not need to be taken into account.}
\begin{equation}
\begin{split}
S_{\phi}^{(2)} = \frac{1}{2}N(\alpha_q)\int\!d^4y \Big(-(\partial\phi)^2 +\frac{m_{\phi}^2}{y_0^2}\phi^2\Big) \,,
\end{split}
\end{equation}
and thus fluctuations travel at the speed of light. 

Let us now add a source term in the action of the form
\begin{equation}
\begin{split}
S_{J}=\int\!d^4x\, \mathcal{L}_{J}= \frac 1L \int\!d^4x\, e^{-q(x^\mu)/L} J(x^\mu)=\int\!d^4x\, \chi(x^\mu) J(x^\mu)\,.
\end{split}
\end{equation}
In the Weyl representation the source is rewritten as 
\begin{equation}
\begin{split}
S_{J}=\int\!d^4y\,\left|{\rm Det}\frac{\partial x^\mu}{\partial y^\nu}\right| \phi(y^\mu)\left(1+L^2\Omega^2\right) J(y^\mu-L^2e^\pi \left(1+L^2\Omega^2\right)^{-1}\Omega^\mu)\,.
\end{split}
\end{equation}
Expanding at first order in the fluctuations of $\phi$, after a lengthy but straightforward computation, modulo an irrelevant overall constant term,  we get 
\begin{equation}
\begin{split}
S_{J}=\int\!d^4y\,\phi\left(y^0,y^i\right) J\Big( y^0(1-\alpha_q^2)^{-1/2},y^i \Big)\,.
\end{split}
\end{equation}
Thus, the field is coupled non-locally with the source exactly by the amount expected for the fluctuations to effectively travel with subluminal speed $c_s=\sqrt{1-\alpha_q^2}$.

%{}~%%%%%%%%%%%%%%%%%%%%%%%%%%%%%%%%%%%%%
\section{Summary and Conclusions}
\label{sec:con}

In this paper we have studied two non-linear realizations of the Galileon group Gal($3+1$,1). These realizations are the contraction of the two known non-linear realizations of the conformal group, denoted DBI and Weyl representations in \cite{Creminelli:2013fxa}. The two representations are related by a field redefinition that crucially involves the spacetime coordinates as well. These field redefinitions (keeping the spacetime coordinates fixed) contain an infinite number of higher derivative terms. In the regime in which the series can be truncated to a certain number of terms, such as perturbative 
scattering processes involving a finite number of fields in a trivial background, nothing interesting happens: the two theories are simply equivalent.
However, in discussing the propagation of perturbations around a background the series cannot be truncated. The propagation appears qualitatively different in the two representations. How is it possible for example that, starting from a free theory, one can induce propagation outside the Minkowski lightcone using a field redefinition? First of all, we showed that the two representations give the same result if one is interested in a localized background and in its asymptotic effect on a signal. This is not surprising since field redefinitions change the interactions but should not affect asymptotic properties, similarly to what happens to the S-matrix. However, if one is interested 
in local measurements in the presence of a background, the two representations still seem to give different answers about the superluminality with respect to the Minkowski lightcone. The point is that for this kind of measurements one must specify how fields couple with external sources and it turns out that local couplings in one representation become non-local in the other.

Superluminality, starting from \cite{Adams:2006sv}, has been used as a way to rule out a conventional, local and Lorentz invariant UV completion. 
It is important to emphasize that an underlying assumption is that causality is defined by the Minkowski lightcone.
This is typically taken for granted (and implicitly assumed in \cite{Adams:2006sv}), and is the reason why  we  insisted in keeping the external sources fixed (or equivalently the coordinates fixed). 
In this context, if one studies the Lagrangian for the field $q$ obtained from the free field theory without referring  to the map, one can couple it in a generic (and local) way to other particles. In particular, one assumes that there can be other sectors of the theory which  always propagate on the Minkowski lightcone,  independently of any background. Clearly, this breaks the physical  equivalence of the two representations: while in the initial free theory different massless particles move with the same speed, in the second theory different massless particles have different speed.
With these assumptions, the $q$ theory is pathological since an asymptotic superluminality can be generated: even though the integrated effect for $q$ perturbations vanishes, one can obtain an overall effect using, in the regions where the $q$ speed is subluminal, other fields that propagate on the Minkowski lightcone. 

It is worth stressing an alternative procedure.\footnote{We thank C.~de Rham and A.~Tolley for discussions about this point and for sharing their work \cite{deRham:2014lqa} on the subject.} Instead of keeping the sources fixed, one could perform a redefinition of the source $J$ in eq.~(\ref{LJsource}):
\be
\int d^4y \, \pi(y) J(y) = \int d^4x\, {\rm Jac}(q(x))\, \Big(q(x) -\frac 1{2\Lambda^3} (\partial q)^2 \Big) \tilde J(x)\,,
\ee
where $\widetilde J(x) = J(y)$. In this way one does not induce any non-locality: we simply look at the change of coordinates (\ref{eq.coordtransf0}) as a diffeomorphism acting on all the fields of the theory. 
Interactions between all the fields  and $q$ are induced in the mapped theory and as a consequence all the particles
move on the same lightcone, the transformed of the Minkowski lightcone in the $\pi$ representation. In a general background for $q$, no particle is allowed to propagate on the Minkowski lightcone; they all propagate on an ``effective metric'' which depends on $q$ similarly to what happens in General Relativity and hence the Minkowski lightcone no longer defines causality.
In this case it is meaningless to talk about local superluminality with respect to the Minkowski lightcone.
We have simply rewritten the $\pi$ theory in a complicated way: in particular there is no asymptotic superluminality for any field.

We conclude that, without further specifying the couplings with other fields and external sources, the presence of local superluminality does not imply pathologies, as shown by the example of the free theory mapped in the other representation. A less trivial example can be found by considering the conformal Galileons: starting from the Nambu-Goto action in the DBI representation (which is an interacting theory without pathologies), one can obtain the corresponding Weyl Lagrangian which admits local superluminal effects. On the other hand, asymptotic effects are absent in both representations: {\it only asymptotic superluminality is an unambiguous  sign of pathology.}

We concentrated on the propagation of perturbations, because this is a simple process where the relation between the two representations can be studied consistently without truncating the field redefinition. In this case the necessity of keeping an infinite number of operators, and thus be sensitive to non-locality, arises from having a large background field $\bar q$, which makes the condition of measurable superluminality, $\partial\bar q/\Lambda^2 \gtrsim 1$, compatible with the regime of validity of the EFT, $\partial/\Lambda \ll 1$. 
One may ask whether this has an S-matrix counterpart. In a perturbative computation of S-matrix elements with a given number of legs at any given order in perturbation theory,  only a finite number of terms is relevant, and thus the series can be truncated. In this case the two representations give the same result and there is no sign of non-locality. On the other hand, we expect that the inequivalence of the two representations show up also in 
S-matrix elements in the absence of any background and where the series cannot be truncated, for example in the scattering of waves with large occupation number.
It would be nice to have some concrete example of this.

Our results about the inequivalence of different coset constructions for Gal($3+1$,1)/ISO(3,1) and SO(4,2)/ISO(3,1) should apply to other spacetime symmetries as well. 
It would be interesting to consider additional examples to provide further evidence to this claim, for example in the context of coset constructions at finite density \cite{Nicolis:2013sga} and for fluids \cite{Nicolis:2013lma}. 
 
As we have shown in our previous paper, the DBI and Weyl representations of the conformal group can both be obtained from a change of coordinates in AdS$_5$. 
From the AdS$_5$ point of view, roughly speaking, the first arises in the bulk (a brane in AdS$_5$) while the second arises on the boundary.
It might be interesting to see if the non-local relation between the two representations can shed some light on some aspects of the AdS/CFT correspondence.

%%%%%%%%%%%%%%%%%%%%%%%%%%%%%%%%%%%%%%

\subsection*{Acknowledgements}
We thank  N. Arkani-Hamed, M.~Fasiello, R.~Penco, D.~Pirtskhalava, R.~Rattazzi, R.~Rosen, M.~Simonovi\'c  and especially C.~de Rham, S.~Dubovsky, M.~Mirbabayi, A.~Nicolis, A.~Tolley and G.~Villadoro  for useful discussions.
P.C. acknowledges the support of the IBM Einstein Fellowship. The work of E.T. is supported in part by MIUR-FIRB grant RBFR12H1MW.

\appendix
\section{\label{app:TdR}Map between Galileon operators}

In this appendix we show in detail how the contraction of the coset group SO$(4,2)/$ISO$(3,1)$ to Gal$(3+1,1)/$ISO$(3,1)$ turns the map found in \cite{Creminelli:2013fxa} between the two non-linear realizations (DBI and Weyl) of the conformal group into the Galileon ``duality" of \cite{deRham:2013hsa}.
For the Weyl and DBI Galileons we use the definitions and normalizations of our paper \cite{Creminelli:2013fxa}: the interested reader can find there the explicit form in eq.~(4.1) for the first set of 5 operators and in eq.~(4.5) for the latter. We label them (for $i=1,\dots, 5$) $\mathcal{L}_{\pi i}$ and $\mathcal{L}_{q i}$ respectively.    
We further redefine $\pi \to L \pi$ and $q \to L^2 q$ ($L = 1/ \Lambda$) to have $\pi$ and $q$ with mass dimension one. 

In the case of the coset Gal$(3+1,1)/$ISO$(3,1)$, as discussed in section \ref{sec:TwistAlgebra}, the operators in the two representations have exactly the same form, written in terms of the Goldstone $\pi$ or $q$. They are the standard Galileons and we use the following definitions \cite{Goon:2011qf}:
\begin{equation} \label{eq.galileons}
\begin{split}
\mathcal{L}_{G\pi1}&= \pi/L^3 \ , \\
\mathcal{L}_{G\pi2}&=-\frac{1}{2}(\partial\pi)^2 \ , \\
\mathcal{L}_{G\pi3}&=-\frac{L^3}{2}(\partial\pi)^2[\Pi] \ ,  \\
\mathcal{L}_{G\pi4}& =-\frac{L^{6}}{2}(\partial\pi)^2\left([\Pi]^2-[\Pi^2]\right) \ , \\
\mathcal{L}_{G\pi5}& =-\frac{L^{9}}{2}(\partial\pi)^2\left([\Pi]^3-3[\Pi][\Pi^2]+2[\Pi^3]\right)\ ,
\end{split}
\end{equation}
and the same for $\mathcal{L}_{G q i}$. Here for convenience we defined $\Pi$ the matrix of second derivatives of $\pi$, $\Pi_{\mu\nu}\equiv\partial_\mu\partial_\nu\pi$, and $[\dots]$ is the trace operator. The powers of $L$ in front of the operators are introduced to make them of dimension four. 

The Galileons in eq.~(\ref{eq.galileons}) can be obtained starting from the Weyl Galileons and taking the leading term in the limit $\pi \to 0$:
\begin{equation}
\begin{split}
\mathcal{L}_{G\pi}=\text{Lim}_{\pi\rightarrow0} \, N_W \cdot \mathcal{L}_{\pi},
\end{split}
\end{equation} 
where the diagonal matrix  $N$ simply depends on the relative normalization of the operators,
with our choices $N= {\rm diag}(\frac{1}{4},\frac{1}{2},\frac{1}{2},\frac{1}{2},\frac{1}{2})$. 

The limit in the case of DBI Galileons is less direct because each operator now gives a linear combination of  the standard Galileons:\footnote{More precisely, the limit that we take is $q \sim \epsilon^{3/2},$ $qL \sim \epsilon$ for $\epsilon \to 0$.}
\begin{equation}
\begin{split}
\mathcal{L}_{G q}= \text{Lim}_{q \rightarrow0} \, T \cdot \mathcal{L}_{q} \,,
\end{split}
\end{equation}
where
\begin{equation}
\begin{split}
 T = \, \left(
\begin{array}{ccccc}
 \frac{1}{4} & 0 & 0 & 0 & 0 \\
 -1 & 1 & 0 & 0 & 0 \\
 2 & -6 & 1 & 0 & 0 \\
 0 & 24 & -6 & -1 & 0 \\
 0 & -48 & \frac{21}{2} & 4 & -\frac{1}{2} \\ 
 \end{array} 
\right) \, .
\end{split}
\end{equation}
It is clear at this point that when the limit is taken on both sides, the map corresponding to the twist of the conformal algebra derived in \cite{Creminelli:2013fxa} 
\begin{equation}\label{confmap}
\begin{split}
M_{\mathcal{L}_{q} \rightarrow \mathcal{L}_{\pi}}=
\left(
\begin{array}{ccccc}
 0 & \frac{1}{2} & \frac{7}{64} & -\frac{1}{24} &
   -\frac{1}{192} \\
 0 & 0 & -\frac{1}{16} & -\frac{1}{12} &
 -  \frac{1}{48} \\
 4 & 0 & -\frac{11}{8} & 0 & -\frac{1}{8} \\
 0 & 0 & -\frac{3}{2} & 2 & -\frac{1}{2} \\
 0 & -96 & 21 & 8 & -1 \\
\end{array}
\right)
\end{split}
\end{equation}
becomes a transformation that maps the standard Galileons into themselves:
\begin{equation}\label{duality}
\mathcal{L}_{G \pi} = N \cdot M \cdot T^{-1} \cdot \mathcal{L}_{G q} \,.
\end{equation}
There is one last step that has to be done: the fields $\pi$ and $q$ in the Galileon coset are $-2$ times the ones used in the conformal coset defined in \cite{Creminelli:2013fxa}. This is the consequence of the normalization factor $-1/2$ in the definition of the generators $C$ and $\hat C$ in eqs.~(\ref{dilation1}) and (\ref{dilation2}). 
Including this extra factor the transformation in eq.~(\ref{duality}) becomes:
\begin{equation}
\begin{split}
\left(
\begin{array}{ccccc}
 1 & -1 & \frac{1}{2} & -\frac{1}{6} & \frac{1}{24} \\
 0 & 1 & -1 & \frac{1}{2} & -\frac{1}{6} \\
 0 & 0 & 1 & -1 & \frac{1}{2} \\
 0 & 0 & 0 & 1 & -1 \\
 0 & 0 & 0 & 0 & 1 \\
\end{array}
\right)\,.
\end{split}
\label{DualityMap}
\end{equation}
This is exactly the ``duality" obtained in \cite{deRham:2013hsa} when the differences in the conventions are taken into account: $q_{\rm here} = - q_{\rm there}$ and $\mathcal{L}^{\rm here}_{G\pi} = {\rm diag}(1,-12,-6,-4,-5) \cdot \mathcal{L}^{\rm there}_{G\pi}$.

\section{\label{app:finite}Superluminality with a finite field redefinition}
In this appendix we study the case in which the field redefinition is truncated and not composed by an infinite number of terms.
We start again from the Lagrangian for a free particle and use the finite field redefinition (without changing coordinates)
\begin{equation}
\begin{split}
\pi=q+\frac{1}{2\Lambda^3} (\partial q)^2 \;,
\end{split}
\end{equation}
which corresponds to the first term of eq.~(\ref{eq.fieldred}). %[eq.~\eqref{eq.coordtransf}].
The Lagrangian then reads
\begin{equation}	\label{eq.fasmapTrun}
\begin{split}
\mathcal{L}_{G\pi2}\longrightarrow \mathcal{L}= -\frac{1}{2}(\partial q)^2 +\frac{1}{2 \Lambda^3}\square q (\partial q)^2-\frac{1}{8 \Lambda^6}\partial_\mu (\partial q)^2 \partial^\mu (\partial q)^2 \;.
\end{split}
\end{equation}
Obviously, since the field redefinition is truncated, we do not recover a Galileon Lagrangian and   thus, due to the presence of the quartic operator, the equation of motion for the field $q$ is not of second order. Nevertheless when expanded around a non-trivial background $\bar q$, since the second (DGP-like) term in eq.~(\ref{eq.fasmapTrun}) gives a correction to the speed of propagation which is linear in the background, one expects the quartic term to be negligible (its contribution is ${\cal O}(\bar q^2)$).
If this were true, it would be troublesome: since we are dealing with a finite field redefinition, the notion of local operators remains the same before and after the field redefinition and thus nothing could compensate for the superluminality induced by the DGP operator. Of course this is not the case. \\
The issue with the argument just proposed is that when the superluminality is measurable within the EFT the quartic operator cannot be disregarded.
To understand why this is the case, it is sufficient to consider the equation of motion derived from the Lagrangian (\ref{eq.fasmapTrun}):
\begin{equation} \label{eq.motionbackApp}
\begin{split}
\square q -\frac{1}{\Lambda^3}\square q \,\square q+ \frac{1}{\Lambda^3}\partial^\mu\partial^\nu q \,\partial_\mu\partial_\nu q-\frac{1}{2\Lambda^6}\partial^\mu\left(\square (\partial q)^2 \partial_\mu  q \right)=0 \ ,
\end{split}
\end{equation} 
where the last term (coming from the quartic operator) gives higher derivative terms on the fluctuations of the form
\begin{equation} \label{eq.corrections}
\begin{split}
\frac{\partial \bar q}{\Lambda^3}\frac{\partial \bar q}{\Lambda^3} \square^2q \ , \qquad \frac{\partial \bar q}{\Lambda^3}\frac{\partial^2 \bar q}{\Lambda^3} \partial \square q \;.
\end{split}
\end{equation}
As discussed above, see eq.~\eqref{eq:measurable}, the superluminality induced by the DGP operator is measurable if 
\begin{equation}\label{eq:measure}
\frac{1}{\Lambda^3} \partial \bar q \gtrsim \omega^{-1} \;.
\end{equation} 
When this happens the first term of eq.~\eqref{eq.corrections} becomes of order one and cannot be neglected:
\begin{equation}
\begin{split}
 \frac{\partial \bar q}{\Lambda^3}\frac{\partial \bar q}{\Lambda^3} \square^2q \sim \frac{\square}{\omega^2} \square q \sim \square q \ . 
\end{split}
\end{equation}

The measurability of the effect within the EFT makes sense if the scale $\Lambda$ is interpreted as the cutoff which suppresses all the higher-dimension operators, while of course here we are just looking at a free theory in disguise. However, even if we forget about the EFT point of view and we simply care about the solutions of the differential equation~\eqref{eq.corrections}, we reach the same conclusion: one cannot neglect the quartic operator when talking about superluminality around weak backgrounds (recall also the discussion below eq.~(\ref{eq.wave})).

Notice that the situation with Galileons is very different.  Since in the equation of motion there are two derivatives per field, the quartic Galileon simply corrects the speed of propagation
\begin{equation}
\begin{split}
\delta c_s^{(2)}\sim \left(\frac{\partial^2 \bar q}{\Lambda^3}\right)^2\ ,
\end{split}
\end{equation}
which is indeed smaller than the effect due to the DGP term for $\partial^2 \bar q/\Lambda^3 \ll 1$.

%%%%%%%%%%%%%%%%%%%%%%%%%%%%%%%%%%%%

\end{document}